\documentclass[12pt,preprint]{aastex}

\slugcomment{Not to appear in Nonlearned J., 45.}

\shorttitle{Phosphorous chemistry in the shocked region L1157 B1}
\shortauthors{Aota $\&$ Aikawa}

\begin{document}


\title{Phosphorous chemistry in the shocked region L1157 B1}


\author{T. Aota and Y. Aikawa}
\affil{Department of Earth and Planetary Sciences, Kobe University, Kobe 657-8501,Japan}







\begin{abstract}
We study the evolution of phosphorous-bearing species in one-dimensional C-shock models.
We find that the abundances of P-bearing species depend sensitively on the elemental abundance of P in the
gas phase and on the abundance of N atoms in the pre-shock gas.
The observed abundance of PN and the non-detection of PO towards L1157 B1 are
reproduced in C-shock models with shock velocity $v=20$ km s$^{-1}$ and pre-shock
density {\it n}(H$_2)= 10^4-10^5$ cm$^{-3}$, if the elemental abundance of P in the gas phase is
$\sim 10^{-9}$ and the N-atom abundance is {\it n}(N)/{\it n$_{{\rm H}}$} $ \sim 10^{-5}$
in the pre-shock gas.
We also find that P-chemistry is sensitive to O- and N-chemistry, because N atoms are destroyed mainly by OH and NO.
We identify the reactions of O-bearing and N-bearing species that significantly affect P chemistry.

\end{abstract}





\section{Introduction}
Although the solar abundance of phosphorus is $3\times10^{-7}$ with respect  to H (Asplund et al. 2009),
only a small number phosphor-containing molecules have been observed in space.
PN is the only P-bearing species detected in 
Turner \& Bally (1987) detected PN in Ori(KL), W51M, and Sgr B2, whereas  Matthews, Feldman \& Bernath (1987) did not detect PO in Ori(KL) or Sgr B2.
Considering the non-detection of PO, Turner \& Bally (1987) suggested that PN
is formed by grain disruption.
On the other hand, Millar et al. (1987) calculated the P-chemistry considering related reactions
(e.g., PO + N $\to$ PN + O), and showed that the observed PN abundance and non-detection (i.e., the upper limit) of PO
can be reproduced by gas-phase chemistry at low temperature (50 K).
Later, Charnley \& Millar (1994) showed that PN becomes abundant within $10^4$ yr in hot core models;
assuming that P is initially in the form of PH$_3$ ice, they calculated the gas-phase chemistry after
the evaporation of PH$_3$.

Recently, Yamaguchi et al. (2011) detected PN in a low-mass star-forming region, L1157 B1, for the first time.
L1157 is a dark cloud harboring a Class 0 low-mass protostar, which drives a well-collimated molecular outflow with the dynamical age of
${\rm 1.8 \times 10^{4}}$ yr (Umemoto et al. 1992).
L1157 B1 is a shocked region formed by interactions between the blue-shifted molecular outflow and ambient gases. 
Since B1 is distant from the protostar, the ``pure'' shock chemistry can be investigated.
Many observational studies of the B1 region have been conducted to investigate physical and chemical conditions 
(e.g., Bachiller \& P$\acute{\rm e}$rez Guti$\acute{\rm e}$rrez 1997; Hirano \& Taniguchi 2001; Sugimura et al. 2011).
The kinetic temperature is estimated to be $50 - 170$ K (Hirano \& Taniguchi 2001), and H$_{2}$ density is in the
range of 10$^{4} $ cm$^{-3}$ (Hirano \& Taniguchi 2001) to 10$^{5}$ cm$^{-3}$ (Bachiller \& P${\rm \acute{e} }$rez
Guti${\rm \acute{e}}$rrez 1997) in the  shocked region. 
Previous work concluded that C-shock occurs at B1 (Gusdorf et al. 2008; Viti et al. 2011).
PN is detected also towards the shocked region L1157 B2.
The abundance relative to H$_{2}$ is estimated to be {\it n}(PN)/{\it n}(H$_{2}$) $\simeq (2-6) \times 10^{-10}$ towards B1 and
$(3-7)\times 10^{-10}$ towards B2 (Yamaguchi et al. 2011).
The PN line is blue-shifted, and the line width is as broad as 3.8 km s$^{-1}$.
In addition, PN emission is {\it not} detected toward the protostar position. 
Yamaguchi et al. (2011) thus concluded that PN is formed by shock.
A subsequent work (Yamaguchi et al., in prep) reported that PO is {\it not} detected at B1;
the upper limit of PO abundance relative to H$_{2}$ is $2.5 \times 10^{-10}$.
Detailed modeling of phosphorous chemistry in shocks and comparison of the results with these observations will improve
our understanding of P chemistry.

In this work, we study the evolution of the P-bearing species in interstellar shocks. 
We solve the chemical reaction network along one-dimensional (1D) C-shock models, which mimic the L1157 B1 region.
In \S 2, we describe the physical and chemical model. 
Results of the numerical calculation are presented in \S 3. 
In \S 4, we discuss the difference between our model and that of Charnley \& Millar (1994).
The dependences of P-chemistry on shock parameters are also investigated.
We summarize our conclusion in \S 5.

\section{Model}
\begin{figure}
\epsscale{.80}
\plotone{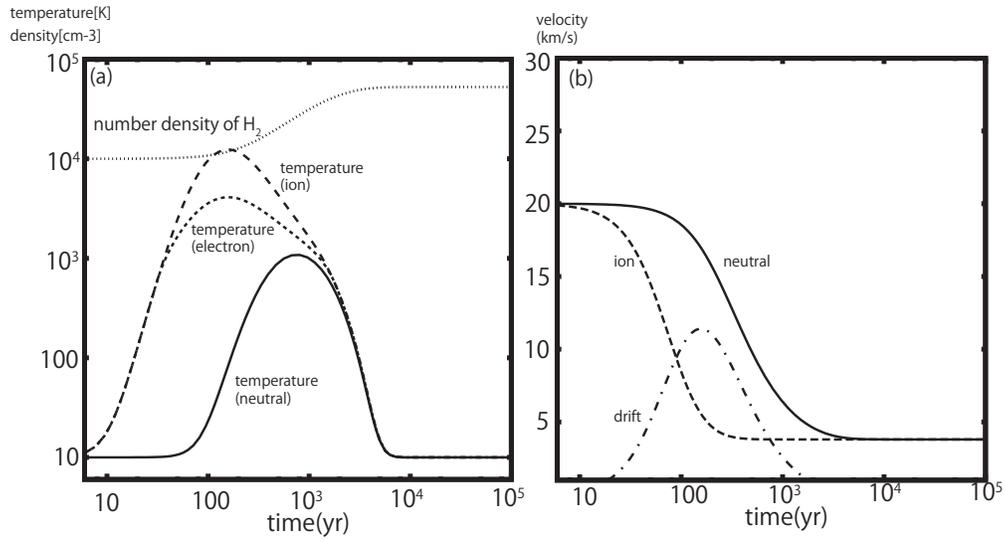}
\caption{(a) Temporal variation of neutral, ion, and electron temperatures and H$_{2}$ density in the C-shock model
with pre-shock density $n$({\rm H$_{2}$}) $=1.0 \times 10^4$ cm$^{-3}$ and velocity $v=20$ km s$^{-1}$.
(b) Temporal variation of the neutral and ion velocities in the C-shock model. Velocity is in the frame co-moving with the
shock front. Dot-dashed line depicts the ion-neutral drift speed ($| u_{ {\rm i} } - u_{ {\rm n} } |$).}
\label{phys}
\end{figure}

We adopt the steady-state 1D C-shock model by ${\rm Jim\acute{e}nez}$-Serra et al. (2008), in which
temporal variations of density and temperature (neutral and ion) are given for several pre-shock densities and velocities.
We make a fitting function of electron temperature, which is not included in the model of ${\rm Jim\acute{e}nez}$-Serra et al. (2008),
based on Draine, Roberge \& Dalgarno (1983) and Flower et al. (1996).
Referring to the observation of L1157 B1 (Bachiller $\&$ P${\rm \acute{e}}$rez Guti${\rm \acute{e} }$rrez 1997), 
we assume the pre-shock density {\it n}(H$_{2})=1.0 \times 10^{4}$ cm$^{-3}$ and velocity $v=20$ km s$^{-1}$.
Figure \ref{phys} shows the temporal variation of neutral, ion, and electron temperatures, H$_{2}$ density, and neutral and ion velocities.
We define $t=0$ as the time when the density and temperature start to rise due to the shock.
The gas is heated to $T\sim 1000$ K in $10^3$ yr, but then cooled to 10 K in $10^4$ yr.
The gas temperature estimated from the observation, 50-170K (Hirano \& Taniguchi 2001), is much lower than the peak temperature of the shock model 
because of the beam dilution; the beam size of Hirano \& Taniguchi (2001) is $\sim$ 0.03 pc, whereas the width of the layer heated to
$T > 100$K is $\sim$ 0.018 pc in the model.

Along the flow, we calculate the chemical reaction network, which consists of 658 species and 11285 reactions.
Chemical reactions and rate coefficients are adopted mainly from the OSU network
(http://www.physics.ohio-state.edu/\verb|~|eric); we use the network of Garrod \& Herbst (2006) at $T\lesssim 100$K
and Harada et al. (2010) at $T>100$K.
We also include collisional dissociations in Table A1 of Willacy et al. (1998),
as well as the reactions of PH$_3$ and PH$_4^+$ in Table 1 of Charnley \& Millar (1994).
Grain surface reactions are included in our chemical network to set up the initial abundances of pre-shock matter (see Model B described below).
In the shock chemistry, however, we neglect the grain-surface reactions, because
they are not effective in the shocked region within the dynamical time scale of L1157 (1.8$\times 10^{4}$ yr).

In C-shock, the neutral and ion temperatures are different. To evaluate the reaction rate coefficient in such flows,
Flower et al. (1985) introduced effective temperature, $T_{ {\rm eff} }$, including the streaming effect:
\[  \frac{3}{2} k_{ {\rm bol} } T_{ {\rm eff} } 
= \frac{3}{2} k_{ {\rm bol} } T_{ {\rm r} } + \frac{1}{2} m_{ {\rm in} } (u_{ {\rm i} } - u_{ {\rm n} })^{2}  \]
\[ T_{ {\rm r} } = \frac{ m_{ {\rm i} } T_{ {\rm n} } + m_{ {\rm n} } T_{ {\rm i} } }{ n_{ {\rm i} } + m_{ {\rm n} } }  \]
\[ m_{ {\rm in} } = \frac{ m_{ {\rm n} } m_{ {\rm i} } }{ m_{ {\rm n} } + m_{ {\rm i} } }  \quad , \]
where $ u_{ {\rm n} }$ and $u_{ {\rm i} }$ are neutral and ion velocities, and $m_{ {\rm n} }$ and $ m_{ {\rm i} }$ are neutral and ion masses, respectively.
We also use the following formula for rate coefficients of collisional dissociation of molecules by ions
\[ k =  \sigma \times | u_{ {\rm i} } - u_{ {\rm n} } | \]
\[ \sigma = 1.0 \times 10^{-16} (1 - \frac{ E_{0} }{ E_{ {\rm T} } } ) \quad {\rm cm^{2}} \quad ,\]
if the kinetic energy of the relative motion of the reactants,
$E_{ {\rm T} }$ ($ = \frac{1}{2} m_{ {\rm in} } | u_{ {\rm i} } - u_{ {\rm n} } |^{2}  $),
is larger than the threshold energy $E_{0}$.

In this work, we consider two models. In model A, we assume that heavy-element species are completely accreted onto grain 
surfaces in the pre-shock gas. In model B, we calculate molecular evolution in a molecular cloud condition to set the 
initial abundance of the pre-shock gas.

In hot core models (e.g., Charnley \& Millar 1994), it is assumed that heavy elements are completely depleted onto grain
surfaces initially due to low temperature (10K) and high density ($n$(H$_2)\sim 10^{6} - 10^{7}$ cm$^{-3}$) in prestellar cores.
Chemical calculation starts when the protostar heats the core and the grain surface species are sublimated.
Referring to the hot core models, we assumed that heavy elements (heavier than He) are completely depleted on
grain surfaces in the pre-shock gas in Model A. We adopt the initial molecular abundances (Table \ref{tab:1}) of
the hot core model of Nomura \& Millar (2004).
They determined these initial abundances to be consistent with the observations of interstellar ices in molecular clouds
and gas-phase species in hot cores.
Previous studies (e.g., Draine, Roberge \& Dalgarno 1983; Caselli, Hartquist \& Havnes 1997) showed that the sputtering of grain-surface species is
effective in C-shock with 20 km s$^{-1}$.
${\rm Jim\acute{e}nez}$-Serra et al. (2008) calculated the sputtering of dust surface species and showed that
the time scale of sputtering is about 100 yr, if the initial H$_{2}$ density is 10$^{4}$ cm$^{-3}$.
Since this sputtering time scale is much shorter than the heating time scale of neutral gas and electrons (Figure \ref{phys}) and
the chemical time scale (see \S 3), we simply start our calculation with all the molecules desorbed to the gas phase at $t=$0.

In model B, we set the initial molecular abundances by calculating the chemical evolution in a molecular cloud condition
({\it n}(H$_{2}$)=10$^{4}$ cm$^{-3}$, $T=10$ K), which is the same as the parameters in our pre-shock gas.
Elemental abundance is the same as Model A.
The molecular cloud calculation starts with atoms and ions except for hydrogen and phosphorus; H is in molecular
form (Table \ref{tab:2}) and P is in the solid phase (see below).
We adopt the molecular abundances at $t=1\times 10^{5}$ yr as the initial condition of the shock chemistry.
At the start of the shock chemistry, all ice-mantle species formed in the molecular cloud are desorbed to the gas phase.

The initial form and abundance of phosphorous relative to hydrogen require some explanation. Although the solar
abundance of phosphorous
is $3\times 10^{-7}$, P-bearing species are not detected in quiescent molecular clouds (e.g., Turner et al. 1990).
If all P is in the gas phase (i.e., P$^+$) initially, theoretical models predict that PN abundance becomes high enough to be easily detected
in molecular clouds. A significant amount of phosphorous thus should be depleted in the solid phase in quiescent
clouds (e.g., Millar et al. 1987), but the dominant form of P in solid is unknown.
Charnley \& Millar (1994) assumed that PH$_3$ ice is as abundant as $1.2\times 10^{-8}$ in the prestellar core,
and that only this amount of phosphorous is desorbed to the gas phase upon heating via star formation.
Following Charnley \& Millar (1994), we assume PH$_3$ abundance of $1.2\times 10^{-8}$ in Model A.

In Model B, we assume that all P is in the solid phase in molecular clouds.
We then assume that a fraction of P is desorbed to the gas phase in atomic form via sputtering in the shock.
To make the comparison with Model A easy, the desorbed abundance of P atoms is set to 1.2$\times 10^{-8}$.
We note that the abundances of P-bearing species are proportional to the assumed elemental abundance
of P (P atom or PH$_3$ initially) in the shocked gas. Although we assume a gaseous P abundance of $1.2\times 10^{-8}$
in our fiducial model, we vary this abundance in order to reproduce the observed abundance of PN and the non detection of PO.
The result tells us how large of a fraction of P is in the relatively volatile component in the solids that is desorbed to the
gas phase via sputtering in the C shock.

\section{Result}

\subsection{Model A} 
\begin{figure}
\epsscale{.60}
\plotone{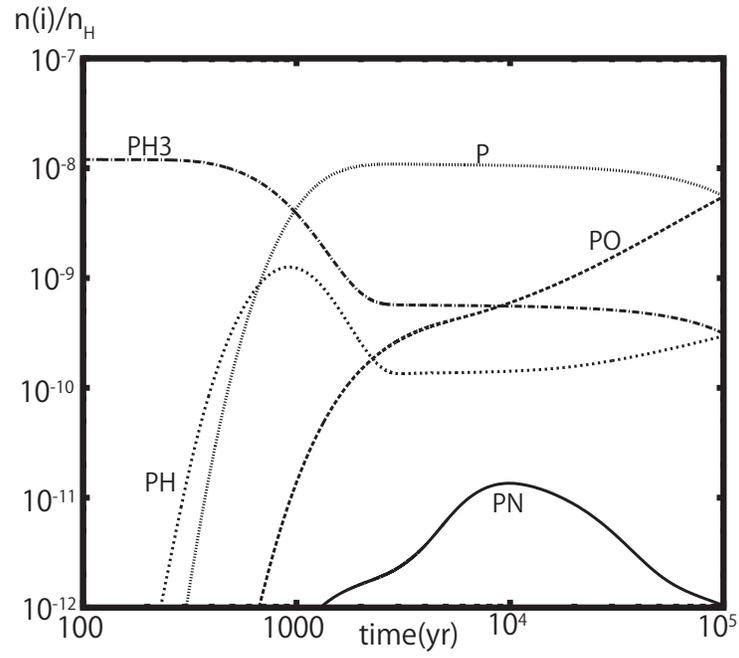}
\caption{Temporal variation of P-bearing species in Model A. Initial N atom abundance is set to zero.}
\label{P-relate-A-0}
\end{figure}

\begin{figure}
\epsscale{.60}
\begin{center}
\fbox{
\plotone{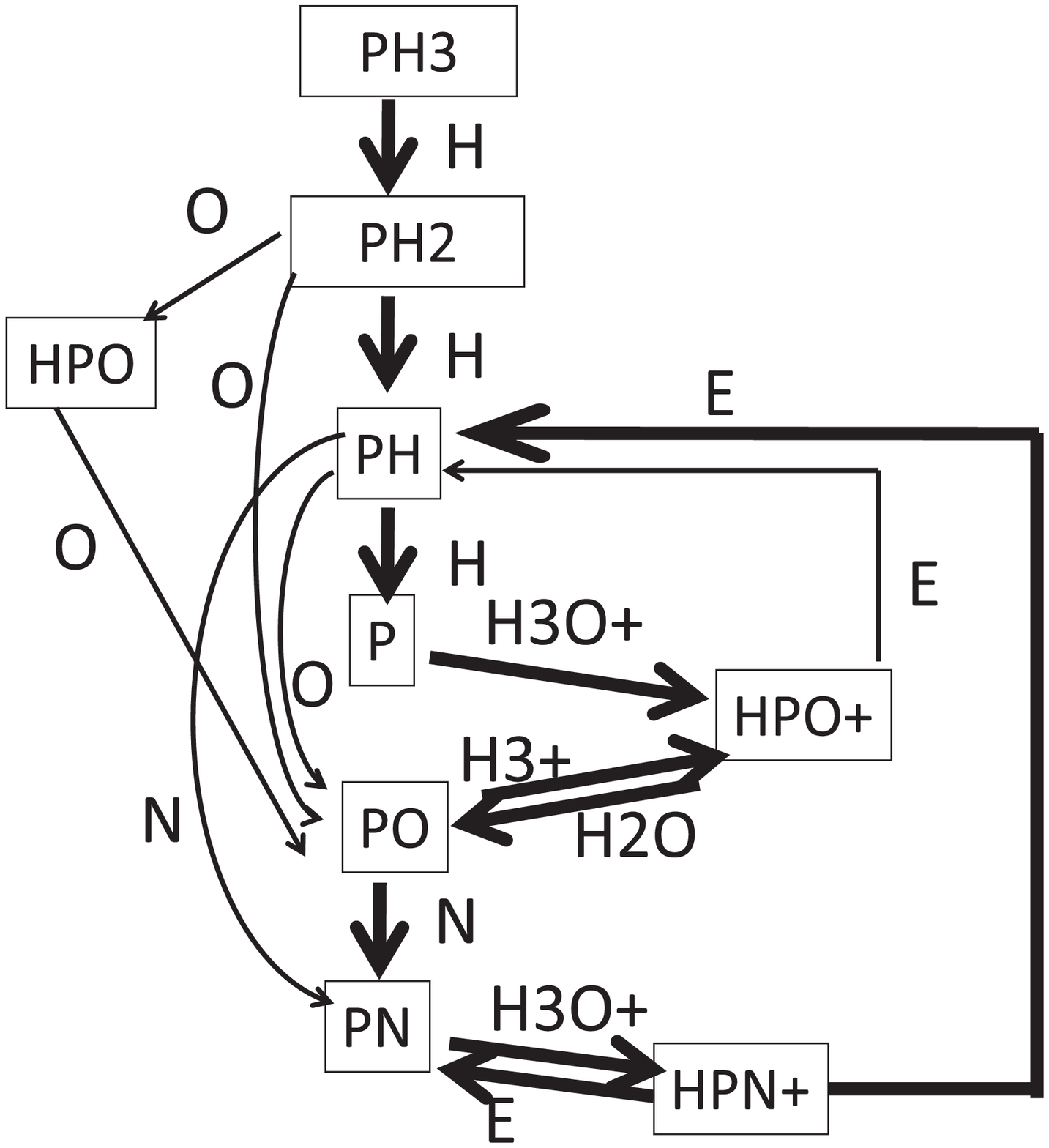}
}
\end{center}
\caption{Reaction network of P-bearing species in the shocked gas.}
\label{P-flow-A}
\end{figure}

Figure \ref{P-relate-A-0} shows the time evolution of P-bearing species in Model A.
Abundances are relative to the number density of hydrogen nuclei ($n_{{\rm H}}$).
Major chemical reactions of P-bearing species are shown in Figure \ref{P-flow-A}.
At first, P is in PH$_{3}$, which reacts with H atoms,  converting them to P atoms:
\begin{equation}
\rm{
  \label{h,ph3-ph2,h2}
  H + PH_{3} \to PH_{2} + H_{2} 
} 
\end{equation}
\begin{equation}
\rm{
  \label{h,ph2-ph,h2}
H + PH_{2} \to PH + H_{2}
}
\end{equation}
\begin{equation}
\rm{
  \label{h,ph-p,h2}
H + PH \to P + H_{2}.
}
\end{equation}
These reactions have activation barriers of 735K, 318K, and 416K, respectively, 
and thus become active only when the gas temperature is higher than $\sim$ 100 K, i.e., from $t=$several $10^2$ yr to
a few $10^{3}$ yr in our model. P atoms react slowly with H$_{3}$O$^{+}$ to form PO
\begin{equation}
\rm{
  \label{p,h3o+-hpo+,h2}
P + H_{3}O^{+}  \to  HPO^{+} + H_{2}
}
\end{equation}
\begin{equation}
\rm{
  \label{hpo+,h2o-po,h3o+}
HPO^{+} + H_{2}O \to PO + H_{3}O^{+} .
}
\end{equation}
PN is then formed by the following reactions
\begin{equation}
\rm{
  \label{po,n-pn,o}
PO + N \to  PN + O
}
\end{equation}
\begin{equation}
\rm{
  \label{ph,n-pn,h}
PH + N \to PN + H
.}
\end{equation}

For comparison with the observation, we refer to the abundances at the dynamical time scale of the outflow of L1157,
$\sim 10^{4}$ yr (Umemoto et al. 1992). In Figure \ref{P-relate-A-0},
PN abundance is lower, and PO abundance is much higher than observed in L1157 B1. Considering 
the reactions (\ref{po,n-pn,o}) and (\ref{ph,n-pn,h}), it is clear that the abundances of PN and PO depend on that of N atoms.
In the current model, the dominant form of nitrogen is N$_2$, and N atoms are formed by the collisional dissociation
of N$_2$
\[ \rm{ N_{2} + Fe^{+} \to N + N + Fe^{+} }   \]
and the following reactions
\[ \rm{  He^{+} + N_{2} \to N^{+} + N + He }  \]
\[ \rm{  N_{2} + cosmic ~ray \to N + N. }  \]
Collisional dissociation is effective only at $t \sim 200$ yr. The latter two reactions proceed slowly and 
thus produce insufficient amounts of N atoms to form PN.
We note here that the initial zero abundance of N atoms would be artificial.
According to theoretical models and observations (Aikawa et al. 2001; Maret et al. 2006),
nitrogen stays in an atomic form for a significant time scale ($\gtrsim 10^5$ yr) in molecular clouds.
Therefore, we next investigate the dependence of PN and PO abundances on the initial abundance of N atoms.

\begin{figure}
\epsscale{1.0}
\plotone{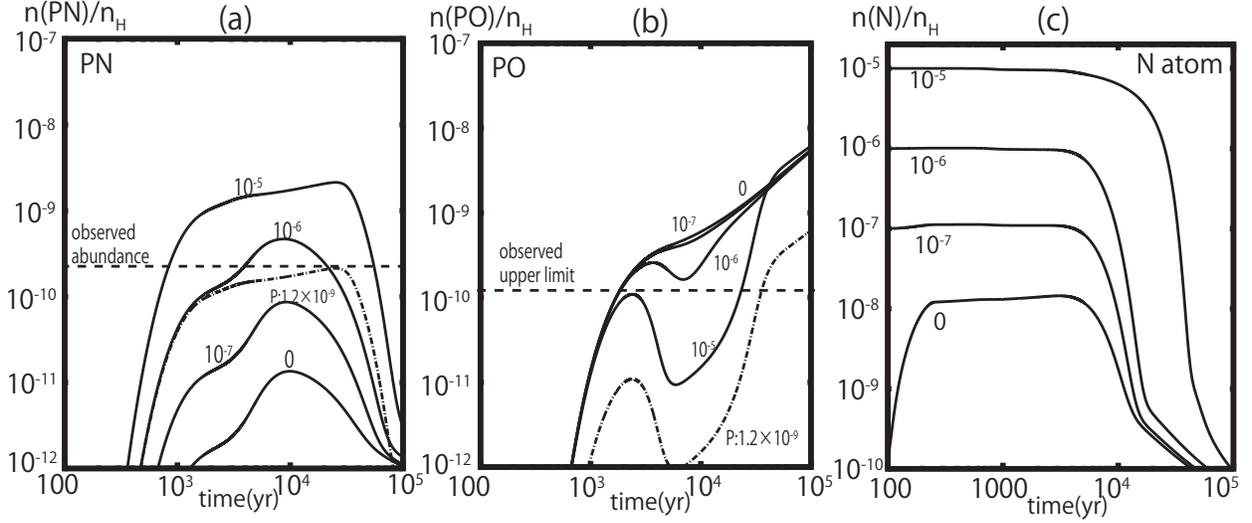}
\caption{PN, PO, and N atom abundances in Model A with various initial N atom abundances, which are each attached to a line.
Dot-dashed lines represent PN and PO abundances with the initial PH$_{3}$ abundance of 1.2$\times 10^{-9}$ and N atom abundance of $10^{-5}$.}
\label{PN-PO-N-comp}
\end{figure}

Figure \ref{PN-PO-N-comp} shows the temporal variation of PN, PO, and N atoms with various initial N atom abundances.
The elemental abundance of nitrogen is kept constant; in a model with high N atom abundance, the initial N$_2$ abundance is low.
As expected, PN abundance is significantly increased, whereas PO abundance is lowered, if the initial N atom abundance is high.
The abundances of other P-bearing species (P atom, PH, and PH$_3$) do not change significantly from those shown in Figure \ref{P-relate-A-0}.
N atom decreases at $t = 10^{4} - 10^{5} $ yr (Figure \ref{PN-PO-N-comp} c);
it is converted to N$_{2}$ by the following reactions 
\begin{equation}
\rm{
  \label{n,oh-no,h}
N + OH \to NO + H
}
\end{equation}
\begin{equation}
\rm{
  \label{n,no-n2,o}
N + NO \to N_2 + O
}
\end{equation}
\[ \rm{  NO + NH_{2} \to N_{2} + H_{2}O. }  \]
Then PN is converted back to PO as N atom decreases (Figure \ref{P-flow-A}).

If the initial N atom abundance is $1.0 \times 10^{-6}$, the observed abundance of PN is
reproduced on a relatively limited time scale $\sim 10^4$ yr, but PO abundance exceeds the observed upper limit.
On the other hand, if the initial N atom abundance is $1.0 \times 10^{-5}$, PO abundance is consistent with the upper limit
for a longer time scale, but
PN abundance slightly exceeds the observed abundance ($(1-3)\times 10^{-10}$ relative to $n_{\rm H}$) at $t \sim 10^4$ yr.
We note here that these abundances of P-bearing species are proportional to the initial PH$_3$ abundance.
In the current calculation, we assumed the initial PH$_{3}$ abundance of $1.2\times 10^{-8}$ following Charnley \& Millar (1994).
However, this assumption is rather arbitrary;
Charnley \& Millar (1994) did not compare their PN abundance with observation quantitatively, because the abundance
estimates in hot cores were uncertain due to beam dilution and difficulty in estimating the H$_2$ column density. 
Thus far, there are no direct observational constraints on PH$_3$ abundance.
Dot-dashed lines in Figure \ref{PN-PO-N-comp} show the temporal variation of PO and PN with the initial
PH$_3$ abundance of $1.2\times 10^{-9}$ and N atom abundance of 10$^{-5}$;
both PN and PO abundances are consistent with the observation.

We also note that the initial form of desorbed phosphorus is not necessarily PH$_{3}$, since there are no observational constraints on P in solids.
Our results do not change if P is desorbed as P atoms, instead of PH$_{3}$, because P atoms are on the path to PN in the
chemical network (see Figure \ref{P-flow-A}).

\subsection{Model B} 
\begin{figure}
\epsscale{.60}
\plotone{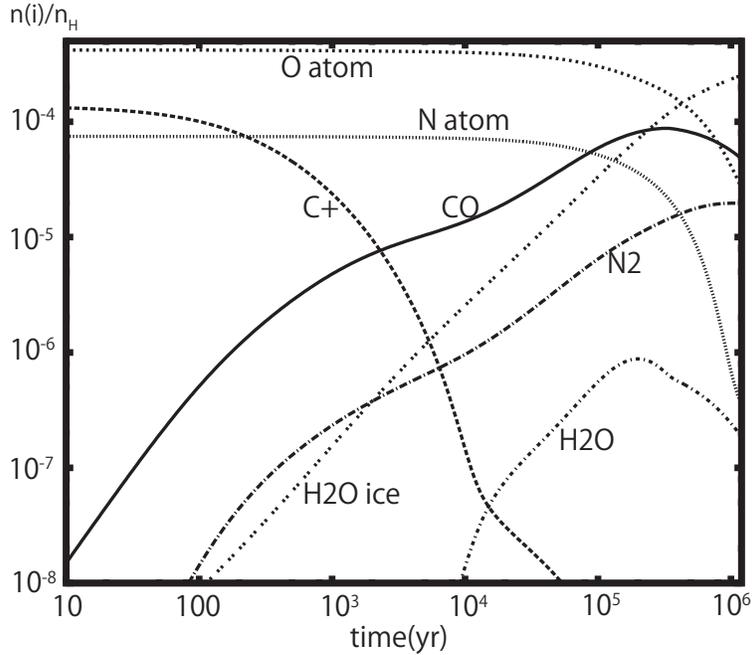}
\caption{Molecular evolution in the molecular cloud condition, $n_{{\rm H}}= 2.0 \times 10^4$ cm$^{-3}$ and $T=10$ K.}
\label{P-relate-m}
\end{figure}

In Model A and Figure \ref{P-flow-A}, we can see that N atoms, O atoms, and protonated water are major reactants with P-bearing species.
This indicates that P chemistry may depend on the initial abundances of water and O atoms, as well as on the N atom abundance.
In order to check the dependence of P chemistry on initial molecular abundances, here we set the initial abundances by calculating molecular evolution in a molecular cloud condition,
$n({{\rm H}_{2}})=1.0 \times 10^{4}$ cm$^{-3}$ and $T=10$ K (Figure \ref{P-relate-m}).
We adopt the molecular abundance at $1\times 10^{5}$ yr as the initial abundance for the shock model.
Ice mantle species are assumed to be desorbed to the gas phase as in Model A.
We assume that phosphorus is totally depleted in the solid phase in molecular clouds, but a fraction ($1.2 \times 10^{-8}$
relative to hydrogen nuclei) of P is desorbed to the gas phase in atomic form via sputtering in the shock.

\begin{figure}
\epsscale{.80}
\plotone{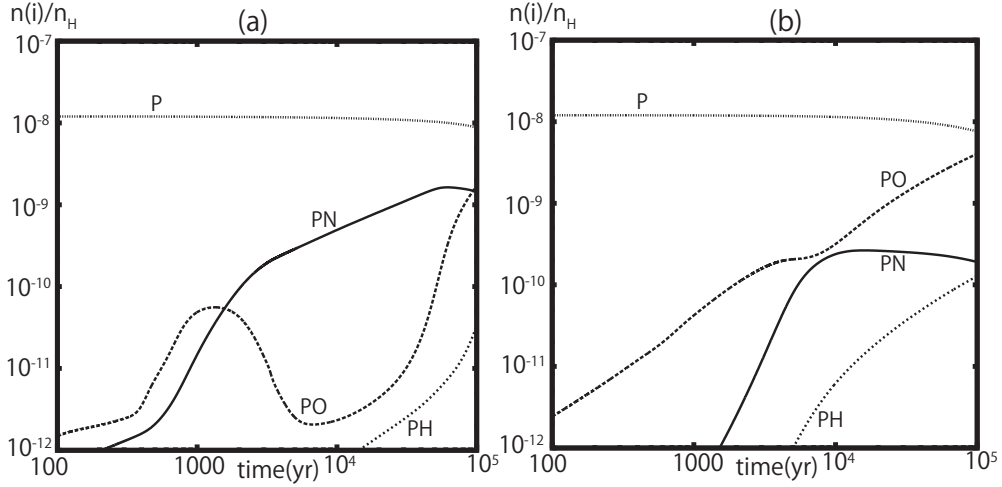}
\caption{Temporal variation of P-bearing species in Model B. We adopt the molecular cloud abundances at
(a) $t=1\times 10^5$ yr and (b) $t=1\times 10^6$ yr as the initial conditions of shock chemistry. P atoms are assumed
to be desorbed to the gas phase by sputtering in the shock.}
\label{P-relate-S}
\end{figure}

Figure \ref{P-relate-S} ($a$) shows the temporal variation of P-bearing species in the C shock model.
Major chemical reactions of P-bearing species are mostly the same as in Model A.
Since N atoms are as abundant as 10$^{-5}$ initially, PN becomes abundant, whereas PO is depleted at $t\sim 10^4$ yr.
If we adopt the abundances at $t=1\times 10^6$ yr in the molecular cloud as initial conditions for the shock chemistry,
PO abundance exceeds the observed upper limit at $t\sim 10^4$ yr (Figure \ref{P-relate-S}b),
because the initial N atom abundance is low ($\sim 7.0 \times 10^{-7}$).

As in Model A, the abundances of P-bearing species are proportional to the assumed elemental abundance of P in the
gas phase. If we adopt the molecular cloud abundance at $10^{5}$ yr and the initial P atom abundance of $\sim 3\times 10^{-9}$, our PN and PO abundances become consistent with the observation at $\sim 10^{4}$ yr.

\section{Discussion}
\subsection{Differences between our results and those of Charnley \& Millar (1994)}
In our model, PN becomes abundant {\it only} if initial N atom abundance is high ($\gtrsim 10^{-5}$).
In the hot core model of Charnley \& Millar (1994), on the other hand, 
abundant PN is formed in $\sim$ 10$^{4}$ yr even though N is all in N$_{2}$ initially.
Why do we need abundant N atoms? Is it due to the difference in physical models (hot core vs. shock) or chemical reaction networks?
We note that Charnley \& Millar (1994) used the network model of UMIST91 (Millar et al. 1991).
Since reaction rate coefficients have been updated significantly since UMIST91, it is worth investigating how our network model differs from UMIST91,
and if our conclusion, that there is high N atom abundance in pre-shock gas, sensitively depends on the reaction network model.

\subsubsection{Comparison between chemical network models}
In order to compare our network model with UMIST91, we calculated the same hot core model as Charnley \& Millar (1994).
We adopted the initial abundance in their Table 2; it is essentially similar to our initial condition of Model A with abundant N$_2$ 
and saturated molecules such as H$_2$O.
Temperature and density are set to {\it T}=100 K and {\it n}{\rm (H$_{2}$)} = $10^{7}$ cm$^{-3}$.

\begin{figure}
\epsscale{.80}
\plotone{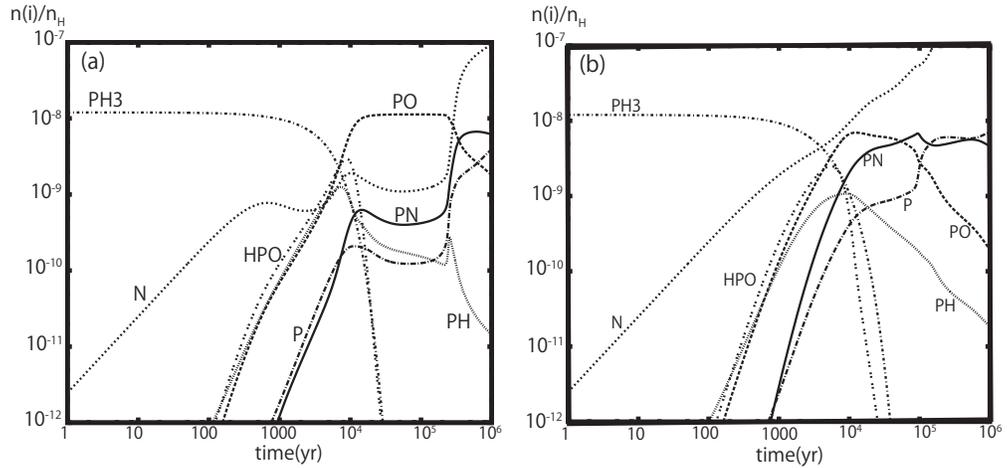}
\caption{Temporal variation of P-bearing species in the hot core model of {\it T}=100K and {\it n}$_{{\rm H}}$=2.0$\times 10^{7}$ cm$^{-3}$.
We use our original network model in (a) and the modified network model in (b).}
\label{P-relate-H}
\end{figure}

The result is shown in Figure \ref{P-relate-H} (a).
At $t \simeq 10^4 -10^5$ yr, PN abundance is much lower, and PO abundance is much higher than that in Charnley \& Millar (1994),
in which PO decreases and PN abundance reaches several $10^{-9}$ at $t\sim 10^4$ yr
(see Figure 1 of Charnley \& Millar 1994).
Nitrogen atoms, which are needed to form PN, are not abundant until $t \sim$ several $10^5$ yr; although N atoms are formed by N$_{2}$ + He$^{+}$,
their abundance is suppressed by reactions (\ref{n,oh-no,h}) and (\ref{n,no-n2,o}).
Although N atom abundance is not shown in Charnely \& Millar (1994), it might be higher than in our model.
Since N atom abundance depends on N-chemistry and O-chemistry (see reactions (\ref{n,oh-no,h}) and (\ref{n,no-n2,o})), 
we compared the rate coefficients of major N- and O- reactions. 
Table 3 lists several reactions that we found to have different rate coefficients in our network and UMIST91.
For example, the rate coefficient of
\begin{equation}
\rm{
  \label{co,oh-co2,h}
CO + OH \to CO_{2} + H
}
\end{equation}
is 500 times higher in UMIST91 than in our model at T=100K;  the rate coefficient in UMIST91 is adopted from Smith
(1998), whereas our rate coefficient is from Frost et al. (1993). 
For reactions that are not included in UMIST91, we put 0 in the column of $A$ parameter of "UMIST91" in Table \ref{tab:3}.
It is natural that most of the reactions listed in Table \ref{tab:3} are
neutral-neutral reactions, whose rate coefficients are hard to measure at low temperatures in laboratories.

We adopted the rate coefficients of UMIST91 for the reactions in Table 3 (hereafter, modified network model) and calculated the hot core chemistry of Charnley \& Millar (1994) (Figure \ref{P-relate-H}b).
Now, the temporal variation of P-bearing species is more similar to that in Charnley \& Millar (1994) than to that in Figure \ref{P-relate-H} (a).
We can conclude that there are differences in N-chemistry and O-chemistry between our network and UMIST91, which significantly affect P-chemistry.

\subsubsection{Shock chemistry with the modified network model}
\begin{figure}
\epsscale{0.90}
\plotone{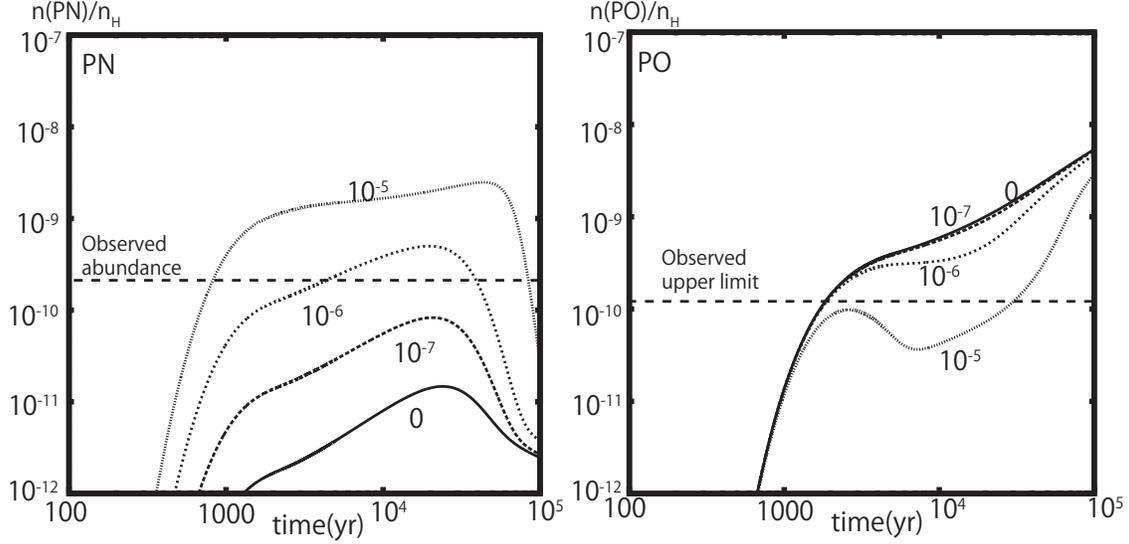}
\caption{PN and PO abundances in the shock model as shown in figure \ref{PN-PO-N-comp} but with the modified network model.}
\label{PN-PO-comp}
\end{figure}

Using the modified network model, we calculated the C-shock model to see how the evolution of P-bearing species differs
from our original model.
The initial abundances are the same as in Model A.
Figure \ref{PN-PO-comp} shows the PN and PO abundances with various initial N atom abundances 
($0,1.0 \times 10^{-7},1.0 \times 10^{-6}$, and  $1.0 \times 10^{-5}$).
Although these results are slightly different from those in Figure \ref{PN-PO-N-comp}, our original conclusion stands: PN and PO abundances become
consistent with the observed value and the upper limit {\it only} if N atom abundance is as abundant as $\gtrsim 10^{-5}$.

\subsubsection{Dependence of hot core model on physical parameters}
Using the modified network model, PN becomes abundant at $\sim 10^4$ yr in the hot core model (as in Charnley \& Millar 1994)
even if the initial N atom abundance is zero. In the shock model, on the other hand, we still need initially high
N atom abundance to make PN abundant at $\sim 10^4$ yr. Since we are now comparing the two models with the same
chemical reaction network, the difference should be attributable to the physical conditions.
In the C-shock model for L1157 B1,
density ({\it n}(H$_2)=10^4$ cm$^{-3}$) is much lower than that in the hot core model ({\it n}(H$_2)=10^7$ cm$^{-3}$), and
the temperature varies temporally from 10 K to 1000 K, whereas the hot core model has a fixed temperature (100 K).
In the following we calculate hot core models with various densities and temperatures in order to determine
density, temperature, or both affect the P chemistry.
We use the modified reaction network and the initial conditions shown in Table 2 in Charnley \& Millar (1994).

\begin{figure}
\epsscale{0.90}
\plotone{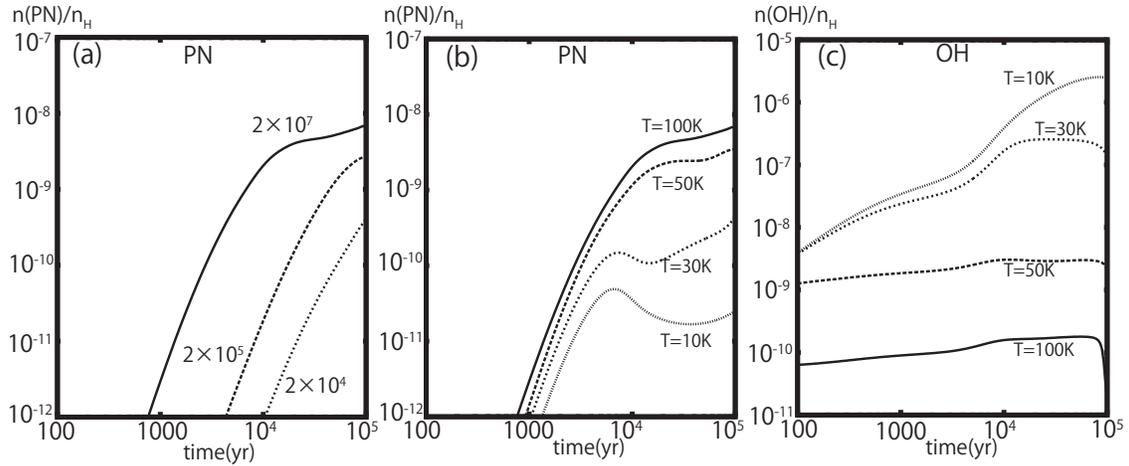}
\caption{ (a) Dependence of PN abundance on gas density
($n_{{\rm H}}$=$2.0 \times 10^{7}$, $2.0 \times 10^{5}$, $2.0 \times 10^{4}$ cm$^{-3}$), and (b, c) dependence of PN and OH abundances
on temperatures ({\it T}=10, 30, 50, 100K) in hot core model.}
\label{PN-OH-comp}
\end{figure}

Figure \ref{PN-OH-comp}(a) shows the PN abundance in hot core models of {\it T} = 100 K with the gas densities of
{\it n}(H$_2)=1.0 \times 10^{4}$ cm$^{-3}$, $1.0 \times 10^{5}$ cm$^{-3}$, and $1.0 \times 10^{7}$ cm$^{-3}$.
We can see that PN is formed in shorter time scales at higher densities. In general, chemical reactions proceed
faster at higher densities. Whereas the collisional time scale of gaseous particles is proportional to $n^{-1}$,
the chemical time scale’s dependence on density is weaker than $n^{-1}$, because the ion-molecule reactions play major roles in the reaction network
and because the ionization degree is proportional to n$^{-1/2}$ .

Figure \ref{PN-OH-comp} (b) shows the PN abundance in hot core models of {\it n}(H$_2)=1.0 \times 10^{7}$ cm$^{-3}$ 
with $T=10$ K, 30 K, 50 K, and 100 K. At lower temperatures, PN is less abundant, because the N atom is destroyed by the reaction
with the OH (reaction(\ref{n,oh-no,h})).
Figure \ref{PN-OH-comp}(c) shows the OH abundance in the same models as in Figure \ref{PN-OH-comp}(b).
OH increases with time at low temperatures (10 K and 30 K), whereas at {\it T} = 100 K, OH is destroyed by
the reaction (\ref{co,oh-co2,h}), which has an activation barrier of 390 K in UMIST91.
We note that in the C-shock model (Figure 1), temperature initially increases to $\sim 1000$ K, but decreases to
$\sim 10$K after several $10^3$ yrs, when the OH abundance increases and hampers the formation of PN.

From these discussions, we can conclude that both high density and high (constant) temperature help to make PN abundant in the hot core model,
whereas in the C-shock model, we need high N atom abundance in the pre-shock gas in order to make
PN abundant at 10$^{4}$ yr.

\subsection{Dependence on shock parameters}
\begin{figure}
\epsscale{.80}
\plotone{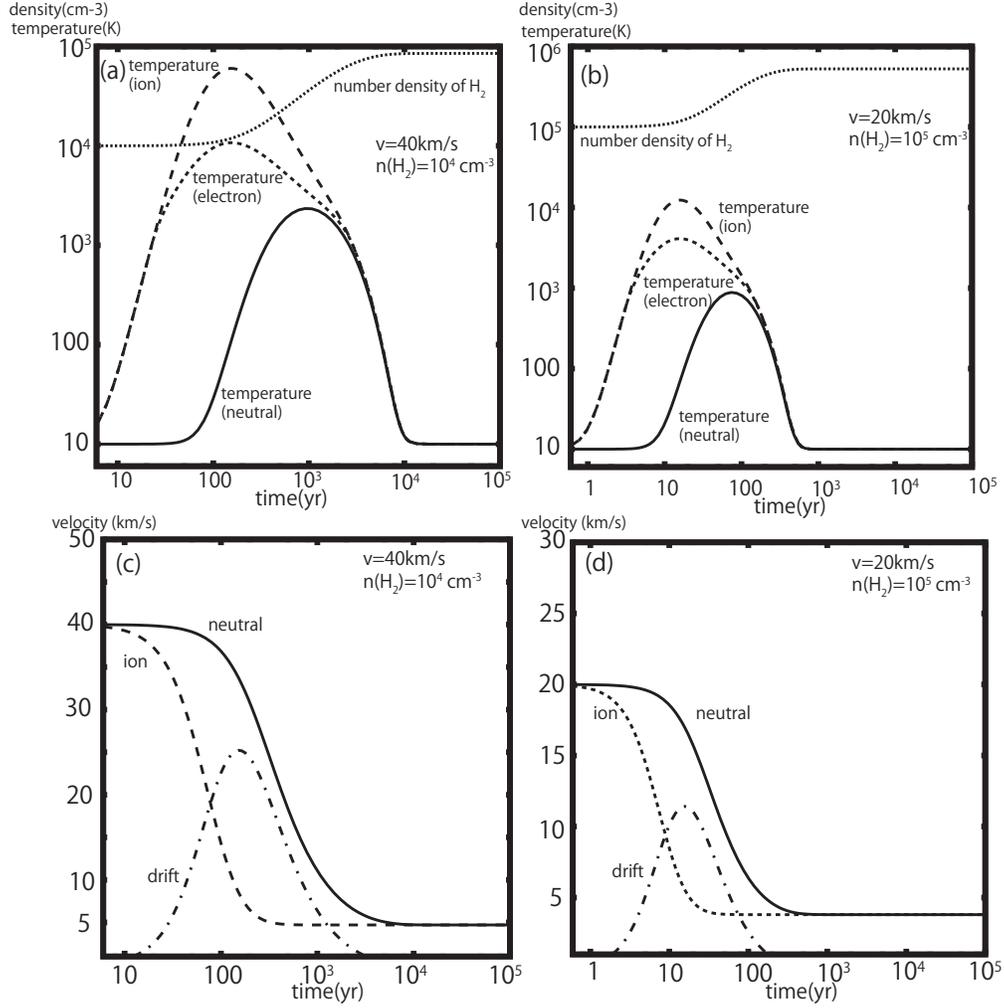}
\caption{Temporal variation of temperature and density in the C-shock model with (a) pre-shock density $n$(H$_{2}$) $=1.0 \times 10^4$ cm$^{-3}$
and  velocity $v=40$ km s$^{-1}$, and (b) $n$(H$_{2}$) $=1.0 \times 10^5$ cm$^{-3}$ and velocity $v=20$ km s$^{-1}$.
(c-d) Temporal variation of the neutral and ion velocities in the two C-shock models.}
\label{phys-diff}
\end{figure}

So far, we have adopted the C-shock model of {\it n}(H$_{2}$)=1.0$\times 10^{4}$ cm$^{-3}$ and $v= 20$ km s$^{-1}$ (Figure \ref{phys}).
In this section, we investigate the dependence of P-chemistry on C-shock parameters: pre-shock density and velocity.
We use our original reaction network and adopt the initial conditions in Model A. The initial N atom abundance is 1.0$\times 10^{-5}$.
Figure \ref{phys-diff} shows the temporal variation of neutral, ion, and electron temperatures, H$_{2}$ density,
and neutral and ion velocities in two C-shock models: (a, c) the pre-shock density
$n({\rm H}_2)=10^4$ cm$^{-3}$ and velocity $v=40$ km s$^{-1}$, and (b, d) $n({\rm H}_2)=10^5$ cm$^{-3}$ and $v=20$ km s$^{-1}$.
The peak temperature is higher in the model with higher velocity, whereas the cooling time is shorter in the model with higher pre-shock density.

\begin{figure}
\epsscale{.80}
\plotone{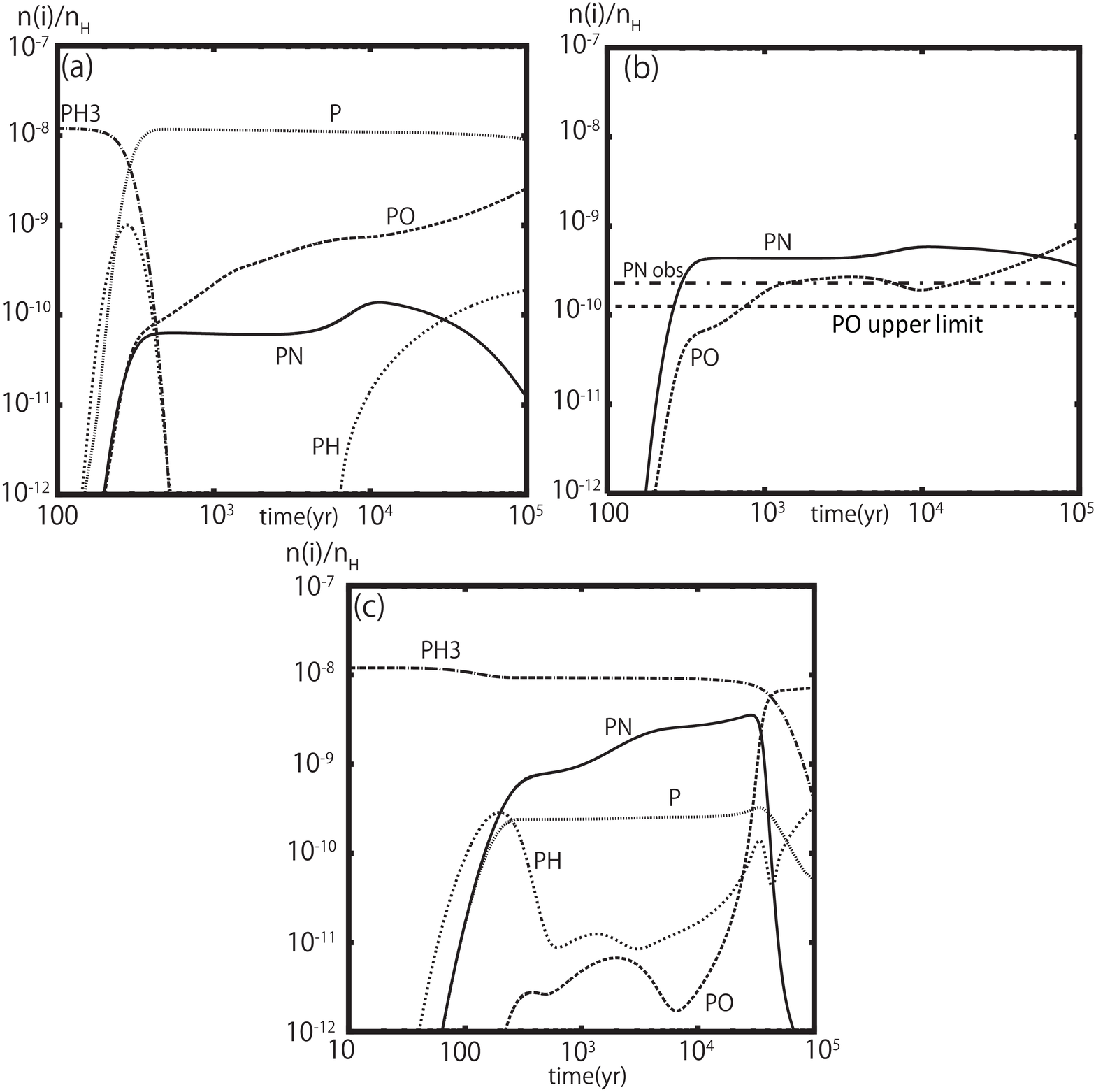}
\caption{Temporal variation of P-bearing species in the C-shock model with pre-shock density $n_{{\rm H}}=2.0 \times 10^4$ cm$^{-3}$ 
and velocity $v=40$ km s$^{-1}$ (a-b), and $n_{{\rm H}}=2.0 \times 10^5$ cm$^{-3}$ $v=20$ km s$^{-1}$ (c).
Initial N atom abundance is $1.0\times 10^{-5}$ in (a) and (c), whereas it is $7.46\times 10^{-5}$ in (b).
Dashed and dot-dashed lines in panel (b) depict the observed PN abundance and upper limits on PO abundance, respectively.}
\label{P-relate-4-40-5-20}
\end{figure}

Figure \ref{P-relate-4-40-5-20} (a) shows the temporal variation of P-bearing species in the model of $n({\rm H}_2)=10^4$ cm$^{-3}$ and
$v=40$ km s$^{-1}$. Since the peak temperature is higher than the model with $v=20$ km s$^{-1}$,
N atoms are depleted in the high-temperature region ($T\sim 2000$K) by
\begin{equation}
\rm{
  \label{h2,n-nh,h}
H_{2} + N \to NH + H,
} 
\end{equation}
which has an energy barrier of $1.66 \times 10^{4}$ K. Then, PN is less abundant and PO is more abundant than Figure \ref{PN-PO-N-comp} (b).
NH is further hydrogenated to NH$_3$ by
\begin{equation}
\rm{
  \label{nh,h2-nh2,h}
NH + H_{2} \to NH_{2} + H
} 
\end{equation}
\begin{equation}
\rm{
  \label{nh2,h2-nh3,h}
NH_{2} + H_{2} \to NH_{3} + H,
} 
\end{equation}
which also have activation barriers.
At high temperatures, H atoms increase due to collisional dissociation of H$_{2}$, which starts the back reactions of
(\ref{h2,n-nh,h})-(\ref{nh2,h2-nh3,h}).
However, back reactions are not efficient enough to stop the decline of N atom abundance.
Figure \ref{P-relate-4-40-5-20} (b) shows the same shock model but with the initial N atom abundance
of 7.46$\times 10^{-5}$; nitrogen is initially all in atomic form. PN is more abundant than PO at $\sim 10^4$ yr.
If we reduce the initial PH$_3$ abundance by a factor of a few, PO abundance can be marginally lower than the
upper limit, whereas PN abundance is in the range of observed values.

Figure \ref{P-relate-4-40-5-20} (c) shows, on the other hand, the temporal variation of P-bearing species with the pre-shock density
{\it n}(H$_{2}$)=1.0$\times 10^{5}$ cm$^{-3}$ and velocity $v= 20$ km s$^{-1}$.
The results are similar to those in Figure \ref{PN-PO-N-comp}; PN is abundant and PO is depleted at $t\sim 10^4$ yr.

In summary, the observed abundances of PN and PO are reproduced in the C-shock models
if the peak temperature is less than $T<2000$ K. In the shock model with higher peak temperatures,
only marginal agreement is found with an extremely high initial N atom abundance.

\section{Conclusions}
In this work, we have calculated the phosphorus chemistry in 1D steady-state C-shock models to examine 
how PN is formed and why PO is not detected in the shocked region L1157 B1. 
We found that PN and PO abundances in our model become consistent with the observation, if the following conditions are satisfied:
\begin{itemize}
\item{a fraction of phosphorus, $\sim 10^{-9}$ relative to hydrogen, is desorbed from the grain
to the gas phase as PH$_3$ or P atom via sputtering in the shock}
\item{initial N atom abundance is high ($\gtrsim 10^{-5}$)}
\item{the peak temperature of the shocked gas is $<$ 2000 K.} 
\end{itemize}

We have also compared our reaction network model with UMIST91, which is adopted in the hot core model of Charnley \& Millar
(1994). Phosphorus chemistry sensitively depends on the abundance of N atoms, which in turn depends on N- and O-chemistry.
We identified several neutral-neutral reactions with different rate coefficients in our network models and UMIST91, and these reactions 
significantly affect the P-chemistry.
We confirmed that our conclusion --- a high abundance of N atoms in the pre-shock gas is needed to make PN abundant --- is valid even if we modify the rate coefficients of these reactions to those of UMIST91. 
Furthermore, our results indicate that, even in hot core models, high N atom abundance is needed
to make PO less abundant than PN at $t\sim 10^4$ yr, if we adopt our updated network instead of UMIST91.

\acknowledgments

We are grateful to NRO survey team, especially to S.Yamamoto and T.Yamaguchi for providing observational results before the publication.
We also thank to V.Wakelam, S.B.Charnley, T.J.Millar, and E. Herbst for helpful comments and discussions.
This work was supported by grant-in-aid for scientific research (23540266, 23103004, 21244021).

\clearpage



\begin{table}[htb]
\begin{center}
\begin{tabular}{|cc|cc|}

\hline
Species & Abundance & Species & Abundance \\
\hline
H$_{2}$ & 0.5 & H$_{2}$S & 1.0$\times$ 10$^{-7}$ \\
He & 0.1 & OCS & 5.0$\times$ 10$^{-8}$ \\
H$_{2}$O & 2.8$\times$ 10$^{-4}$ & Si & 3.6$\times$ 10$^{-8}$ \\
CO & 1.3$\times$ 10$^{-4}$ & PH$_{3}$ & 1.2$\times$ 10$^{-8}$ \\
H & 5.0$\times$ 10$^{-5}$ & S & 5.0$\times$ 10$^{-9}$ \\
N$_{2}$ & 3.7$\times$ 10$^{-5}$ & C$_{2}$H$_{4}$ & 5.0$\times$ 10$^{-9}$ \\
CO$_{2}$ & 3.0$\times$ 10$^{-6}$ & C$_{2}$H$_{5}$OH & 5.0$\times$ 10$^{-9}$ \\
H$_{2}$CO & 2.0$\times$ 10$^{-6}$ & C$_{2}$H$_{6}$ & 5.0$\times$ 10$^{-9}$ \\
O$_{2}$ & 1.0$\times$ 10$^{-6}$ & Fe$^{+}$ & 2.4$\times$ 10$^{-8}$ \\
NH$_{3}$ & 6.0$\times$ 10$^{-7}$ & H$^{+}_{3}$ & 1.0$\times$ 10$^{-9}$ \\
C$_{2}$H$_{2}$ & 5.0$\times$ 10$^{-7}$ & H$^{+}$ & 1.0$\times$ 10$^{-11}$ \\
CH$_{4}$ & 2.0$\times$ 10$^{-7}$ & He$^{+}$ & 2.5$\times$ 10$^{-12}$ \\
CH$_{3}$OH & 2.0$\times$ 10$^{-7}$ & & \\
\hline

\end{tabular}

\caption{Initial molecular abundances relative to hydrogen nuclei in Model A}
\label{tab:1}
\end{center}
\end{table}


\begin{table}[htb]
\begin{center}
\begin{tabular}{|cc|cc|}

\hline
Species & Abundance & Species & Abundance \\
\hline
H$_{2}$ & 5.0$\times$ 10$^{-1}$ & H & 5.0$\times$ 10$^{-5}$ \\
He & 1.0$\times$ 10$^{-1}$ &  S$^{+}$ & 1.55$\times$ 10$^{-7}$ \\
O & 4.18$\times$ 10$^{-4}$ & Si$^{+}$ & 3.6$\times 10^{-8}$ \\
C$^{+}$ & 1.36$\times$ 10$^{-4}$ & Fe$^{+}$ & 2.4$\times$ 10$^{-8}$ \\
N & 7.46$\times$ 10$^{-5}$ & & \\
\hline

\end{tabular}

\caption{Initial abundances in the molecular cloud model for Model B}
\label{tab:2}
\end{center}
\end{table}

\clearpage



\begin{table}[htb]
\begin{flushleft}
\scriptsize
\begin{tabular}{|c|c|c|c|c|c|c|c|c|}
\hline
 & \multicolumn{4}{c|}{OSU} & \multicolumn{4}{c|}{UMIST91}  \\ 
\hline
 reaction & $A$ & $B$ & $C$ & $k(T=100$ K)\tablenotemark{a} & $A$ & $B$ & $C$ & $k(T=100$ K) \\
\hline
CO + OH $\to$ CO$_{2}$ + H & 2.81 (-13)\tablenotemark{b} & 0 & 176 & 4.8 (-14) & 3.1 (-10) & -1.15 & 390 & 2.2 (-11) \\
O + NH$_{2}$ $\to$ NO + H$_{2}$ & 1.0 (-11)& 0 & 0 & 1.0 (-11) & 0 & - & - & -  \\
O + NH$_{2}$ $\to$ NH + OH & 2.0 (-11) & 0 & 0 & 2.0 (-11) & 3.5 (-12) & 0.5 & 0 & 2.02 (-12)  \\
O + NH $\to$ NO + H & 1.16 (-10) & 0 & 0 & 1.16 (-10) & 1.73 (-11) & 0.5 & 0 & 1.0 (-11)  \\
O + PH $\to$ PO + H & 1.0 (-10) & 0 & 0 & 1.0 (-10) & 4.0 (-11) & 0 & 0 & 4.0 (-11)  \\
HPO$^{+}$ + H$_{2}$O $\to$ H$_{3}$O$^{+}$ + PO & --\tablenotemark{c} & --\tablenotemark{c} & --\tablenotemark{c} & 3.9 (-9) & 1.0 (-9) & 0 & 0 & 1.0 (-9)  \\
N + NO $\to$ N$_{2}$ + O & 3.0 (-11) & -0.6 & 0 & 5.8 (-11) & 3.4 (-11) & 0 & 0 & 3.4 (-11)  \\
N + OH $\to$ NO + H & 7.5 (-11) & -0.18 & 0 &  9.14 (-11) & 5.0 (-11) & 0.5 & 0 & 2.89 (-11)  \\
N + CH$_{3}$ $\to$ H$_{2}$CN + H & 8.6 (-11)& 0 & 0 & 8.6 (-11) & 0 & - & - & -  \\
O + OH $\to$ O$_{2}$ + H & 3.5 (-11) & 0 & 0 & 3.5 (-11) & 7.9 (-11) & 0 & 0 & 7.9 (-11)  \\
O + HNO $\to$ OH + NO & 3.8 (-11) & 0 & 0 & 3.8 (-11) & 1.44 (-11) & 0.5 & 0 & 8.31 (-12)  \\
O + NH$_{2}$ $\to$ HNO + H & 8.0 (-11) & 0 & 0 & 8.0 (-11) & 3.2 (-12) & 0 & 0 & 3.2 (-12)  \\
CH$_{5}$O$^{+}$ + e $\to$ CH$_{3}$ + OH + H & 4.64 (-7) & -0.67 & 0 & 9.69 (-7) & 0 & - & - & -  \\
CH$_{5}$O$^{+}$ + e $\to$ CH$_{3}$ + H$_{2}$ O & 8.19 (-8) & -0.67 & 0 & 1.71 (-7) & 0 & - & - & -  \\
CH + H$_{2}$ $\to$ CH$_{3}$ & 3.25 (-17)& -0.6 & 0 & 6.28 (-17) & 0 & - & - & -  \\
CH$_{3}$OCH$^{+}_{4}$ + e $\to$ CH$_{3}$ + CH$_{4}$ + O & 1.5 (-7) & -0.5 & 0 & 2.6 (-7) & 0 & - & - & -  \\
CH$_{3}$OCH$^{+}_{4}$ + e $\to$ CH$_{4}$O + CH$_{3}$ & 1.5 (-7) & -0.5 & 0 & 2.6 (-7) & 0 & - & - & -  \\
O + C$_{3}$H$_{3}$ $\to$ C$_{2}$H$_{2}$ + HCO & 2.31 (-10) & 0 & 0 & 2.31 (-10) & 0 & - & - & -  \\
O + CH$_{2}$ $\to$ HCO + H & 5.01 (-11) & 0 & 0 & 5.01 (-11) & 1.44 (-11) & 0.5 & 2000 & 1.71 (-20)  \\
H + HCO $\to$ H$_{2}$ + CO & 1.5 (-10) & 0 & 0 & 1.5 (-10) & 3.0 (-10) & 0 & 0 & 3.0 (-10)  \\
PO + N $\to$ PN + O & 3.0 (-11) & -0.6 & 0 & 5.8 (-11) & 3.4 (-11) & 0 & 0 & 3.4 (-11) \\
\hline
\end{tabular}
\tablenotetext{a}{Rate coefficient at $T=100$ K.}
\tablenotetext{b}{$a(-b)$ means $a\times 10^{-b}$.}
\tablenotetext{c}{The rate coefficient in OSU model is calculated using equations (3), (5) - (7) of Harada et al. (2010)}
\caption{List of reactions with different rate coefficient in our model and UMIST91 which affect the P-chemistry.
Rate coefficients are given as a function of $A, B$, and $C$: $k=A(T/300)^{B}\exp(-C/T)$.}
\label{tab:3}
\end{flushleft}
\end{table}



\clearpage

\clearpage

\clearpage


\clearpage



\clearpage




\end{document}